\documentclass[journal]{IEEEtran}
\usepackage{xcolor}
\usepackage[pdftex]{graphicx}
\graphicspath{{../pdf/}{../jpeg/}}
\DeclareGraphicsExtensions{.pdf,.jpeg,.png}
\usepackage{graphicx}
\usepackage{hyperref}
\hypersetup{
    colorlinks=true,
    linkcolor=blue,
    filecolor=blue,      
    urlcolor=black,
    citecolor =blue,
    pdftitle={Design of Perfect anomalous reflector},
    pdfpagemode=FullScreen
    }
\usepackage[cmex10]{amsmath}
\usepackage{array}
\usepackage{textcomp}
\usepackage{gensymb}
\usepackage{multirow}
\usepackage{multicol}
\usepackage{stfloats}
\usepackage{amssymb}
\usepackage{cite}

\begin{document}

\title{Efficient Synthesis of 
Passively Loaded Finite Arrays
for Tunable Anomalous Reflection}

\author{
Sravan~K.~R.~Vuyyuru,~\IEEEmembership{Member,~IEEE,}
Risto~Valkonen,~\IEEEmembership{Member,~IEEE,}\\
Sergei~A.~Tretyakov,~\IEEEmembership{Fellow,~IEEE,}
and Do-Hoon~Kwon,~\IEEEmembership{Senior Member,~IEEE}

\thanks{This work was supported in part
  by the European Union’s Horizon 2020 MSCA-ITN-METAWIRELESS project, under the Marie Skłodowska-Curie grant agreement No 956256
  and in part by the US Army Research Office under
  grant W911NF-19-2-0244. \textit{(Corresponding author: Sravan~K.~R.~Vuyyuru.)}}
\thanks{S.~K.~R. Vuyyuru is with 
Nokia Bell Labs, Karakaari 7, 02610 Espoo, Finland
and the Department
 of Electronics and Nanoengineering, School of Electrical
 Engineering, Aalto University, 02150 Espoo, Finland 
 (e-mail: sravan.vuyyuru@nokia.com; sravan.vuyyuru@aalto.fi).}
\thanks{R. Valkonen is with Nokia Bell Labs, Karakaari 7, 02610 Espoo, Finland (e-mail: risto.valkonen@nokia-bell-labs.com).}
\thanks{S.~A. Tretyakov is with the Department of Electronics and Nanoengineering, School of Electrical Engineering, Aalto University, 02150 Espoo, Finland (e-mail: sergei.tretyakov@aalto.fi).}
\thanks{D.-H. Kwon is with the Department of Electrical 
and Computer Engineering, University of Massachusetts Amherst, 
Amherst, MA 01003, USA (e-mail: dhkwon@umass.edu).}
}

\maketitle
\begin{abstract}

A design methodology for planar
loaded antenna
arrays is proposed to synthesize
a perfect anomalous reflection 
into an arbitrary direction
by optimizing the scattering characteristics 
of passively loaded array antennas.
It is based on efficient and
accurate prediction of the induced
current distribution and the associated
scattering for any given set of load impedances.
For a fixed array of finite dimensions,
the deflection angles can be
continuously adjusted with proper tuning of each load. 
We study and develop anomalous reflectors
as semi-finite 
(finite~\texttimes~infinite)
and finite planar
rectangular arrays comprising printed
patches with a subwavelength spacing. 
Anomalous reflection 
into an arbitrary desired angle
using purely reactive loads is 
numerically and experimentally validated. 
Owing to the algebraic nature
of load optimization, the design methodology may
be applied to the synthesis of large-scale
reflectors of practical significance.
\end{abstract}



\section{Introduction}\label{sec:Intro}

\begin{figure*}[t]
\centering
\includegraphics[width=7in,height = 3in]{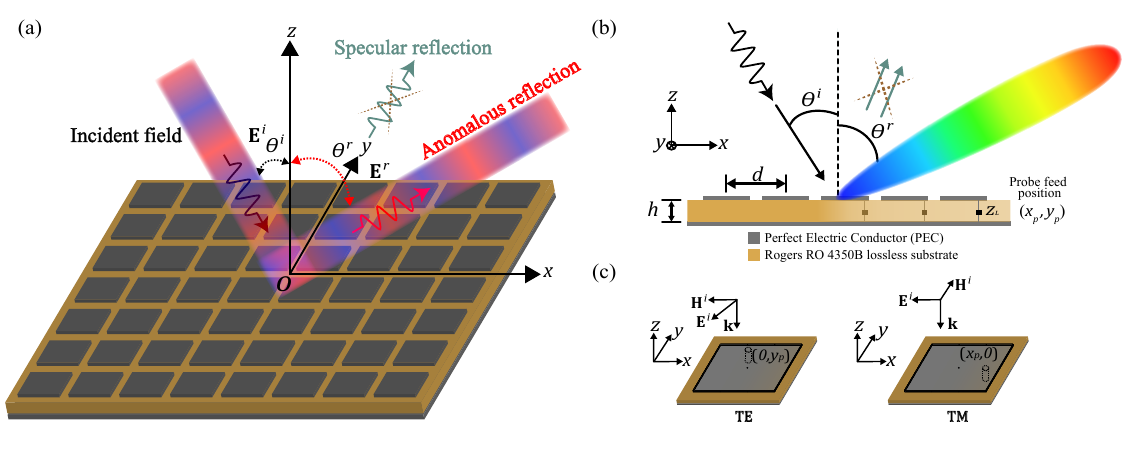}
\caption{(a) Schematic representation of a planar 
array on
a dielectric substrate 
of a thickness $h$ 
backed by
a finite ground plane under a plane wave 
illumination at 
$\theta^i$ and the desired 
large deflection tilt at 
$\theta^r$. 
(b) Side view of the array. The patches are loaded by bulk impedances with an equispaced distance between the patch elements, $d$. Various deflection angles are continuously adjusted by adequately setting purely reactive controllable loads.
The probe feeding point
is at $(x_p,y_p)$. 
The 3-D schematic sketch of a single patch element 
for
(c) transverse electric (TE) and (d) transverse magnetic (TM) waves, where $\mathbf{E}^i$, $\mathbf{H}^i$, and $\mathbf{k}$ are the 
$E$-field, 
$H$-field, and 
the wave vector of 
the incident
wave, respectively.}
\label{fig:figure_theory}
\end{figure*}

\IEEEPARstart{A}{reconfigurable}
Intelligent Surface (RIS) is an engineered 
surface structure
to manipulate electromagnetic (EM) waves 
upon reflection/transmission,
aiming for versatile functionalities~\cite{Smart_Radio_Environments,MarcoCommModelsRIS,Int_RIS_AnaSergei} in 6G wireless networks 
by enhancing coverage 
and overcoming challenges associated with
non-line-of-sight scenarios. In particular, 
an RIS can scatter the incident 
power into a non-specular direction by 
breaking the law of reflection for
a homogeneous reflecting surface, realizing
an anomalous reflection, as
illustrated in Fig.~\ref{fig:figure_theory}. An anomalous reflector encompasses a typically planar surface with 
an
ample amount of
tunable discrete subwavelength-sized elements 
that can present
reconfigurable properties to re-engineer the 
electromagnetic (EM) 
interaction with an incoming wave. Advanced methods to achieve anomalous reflection using periodic metasurfacess~\cite{diaz2017generalized,wong_prx2018,MacroscopicARM2021,kwon_ieeejawpl2018,asadchy2016perfect,vuyyuru,movahediqomi2023comparison} and metagratings~\cite{MG_YounesAlu,casolaro2019dynamic_MG,popov2019Feb,design_MG_LPA_popov,MultipleChannel2020Wang,MG_microwave_popov} have been proposed for various applications. 
Nevertheless,
research on modeling and designing
highly directive and efficient 
anomalous reflectors is still evolving. The most challenging problem is to realize continuous scanning of the reflection angle without deteriorating performance. 

Pioneering research on anomalous deflection has been based on  engineering the reflection phase gradient along 
the reflector surface, 
violating the usual reflection law 
(i.e., the 
incident and reflection angles
are not equal, $\theta^i \neq\theta^r$). 
Application of the phased-array principle \cite{Berry_reflectarray}, also formulated as the generalized law of 
reflection~\cite{yu_science2011},
leads to a periodic reflector with a
linear reflection phase gradient of the local reflection coefficient.
However, the anomalous reflection  efficiency of the
linear
phase-gradient reflectors diminishes gradually 
with increasing
deflection tilts, predominantly due to 
the
impedance mismatch between the incident and 
reflected plane waves and 
the
associated strong 
parasitic reflections in unwanted
directions.

The most typical approach in modeling anomalous reflectors 
treats
periodic structures
because only one period needs to be
analyzed and designed. The spatial periodicity
forces discrete spectra for waves, described by
Floquet modes.
For a given set of incidence and reflection
angles, the smallest spatial periodicity for supporting
them is
$D = \lambda/|\sin\theta^r-\sin\theta^i|$. 
The periodicity changes with the angles,
not allowing continuous scanning for any periodic
reflector with a fixed $D$.
Hence, no periodic reflector
can continuously scan reflected beams.
Traditionally, this
difficulty has been addressed using 
reflectarrays comprising resonant elements
having aperiodic loads of 
a fixed (typically $\lambda/2$) geometrical 
periodicity~\cite{huang2008,nayeri2018}. 
Based on the
locally periodic approximation (LPA), 
array element loads are chosen
by linearly varying the reflection phase of the corresponding uniform arrays 
along the reflection plane.
This technique is effective for
relatively small deflection angles, 
but the
reflection efficiency deteriorates
for wide-angle deflections.
Not relying on the LPA,
design and optimization of the entire reflector
using full-wave EM simulations
is an option, but it
becomes computationally inefficient for electrically 
large structures. 
In addition, a reconfigurable
anomalous reflector design needs to
incorporate tunable
components to manipulate the EM properties 
by external stimuli. 
One approach is
to embed one or more tunable components into each 
meta-atom to enable multi-functional 
reconfigurability~\cite{IntMetaLiu2019}. 

Application of tunable anomalous reflectors
in practice requires synthesis of electrically
large, finite structures incorporating numerous tunable
elements. Even for infinite periodic reflectors
under a supercell periodicity, optimizing the
supercell details for perfect anomalous reflection
using full-wave simulations is a computationally
demanding task.
In alleviating this difficulty,
an arithmetic optimization method has been developed 
using array antenna scattering synthesis for 
infinite periodic 
reflectors comprising reactively loaded patch 
elements in a supercell~\cite{vuyyuru}. 
In this work, we extend the scattering synthesis approach 
to finite antenna arrays
in 2D and 3D,
demonstrating numerically efficient
synthesis of finite reflectors having
high anomalous reflection efficiencies for all deflection angles.

This article is organized
as follows. In Section~\ref{sec:methodology}, we 
present the problem configuration
and develop the scattering synthesis
theory for aperiodically loaded antenna arrays.
Subsections therein
address the principle and optimization technique 
for finite antenna arrays. 
In Section~\ref{sec:finite_Arrays_Opt}, we show 
2-D finite (semi-finite) arrays with
different element spacings,
assess the reflection efficiency, and
evaluate the numerical
efficiency of the optimization. 
In Section~\ref{sec:Planar_Square_Antenna}, we present
optimized finite planar
arrays with aperiodic loadings 
to realize tunable anomalous reflections 
using purely reactive loads. 
In Section~\ref{sec:RISExperimental}, we illustrate the experiment results of the manufactured passive fixed anomalous reflector.
Finally, Section~\ref{sec:conclusion} provides 
a summary and concluding observations.


\section{Principle and Methodology of Anomalous Reflector Design}\label{sec:methodology}

Anomalous reflection into a
predetermined nonspecular direction for a large 
deflection tilt is depicted in 
Figs.~\ref{fig:figure_theory}(a) and (b). 
A finite number of metallic patches
loaded with lumped loads are
etched on a dielectric substrate
as a periodic grid in the $xy$-plane,
at a height $h$ above a finite perfect electric 
conductor (PEC) ground plane.
The plane of incidence and reflection
is the $xz$-plane.
An equispaced grid of patches with 
(generally aperiodic) loads are to be
designed where a
spacing $d$ between radiating elements is 
$d \leq \lambda/2$. With equally spaced radiating elements, 
the traditional reflectarray design
approach uses a reflection
phase distribution that follows 
the linear phase gradient law. 
However,
it is shown in~\cite{vuyyuru} that a perfect 
anomalous reflection in a large deflection angle 
cannot be reached by utilizing a regular loading 
pattern following the linear reflection
phase gradient. Here,
we show that
anomalous finite reflector structures with 
aperiodical loadings can scatter the incoming wave 
from any incidence angle
into a arbitrary non-specular
direction, avoiding computation-intensive full-wave 
EM simulations by transferring the load optimization
to the circuit network synthesis domain.

In an infinite periodic reflector design,
the design goal is to deflect an incoming
propagating plane wave into a reflected
plane wave in an anomalous direction.
For finite arrays subject to a plane-wave
illumination, 
the design goal is to ensure 
one scattered main beam with a
high directivity in the desired direction 
as illustrated in Fig.~\ref{fig:figure_theory}(b). 
A high-efficiency reflection into a
nonspecular angle
entails minimizing
the specular reflection from the underlying PEC 
ground plane for finite structures. 

\subsection{Reception and Scattering Calculation of 
Aperiodically Loaded Antenna}\label{subsec:Re_Scatt}

Let us consider the transmitting and receiving 
configurations of an
antenna, where the reception and scattering 
properties of the loaded antenna depend on the 
load connected at the antenna 
terminals~\cite{balanis2015antenna,stutzman2012antenna}. 
The total scattered $E$-field, $\mathbf{E}^s$, 
scattered by an antenna for a particular load impedance, 
$Z_L$, is the superposition of zero-current 
scattering and port-current 
scattering,
which can be written as~\cite{collinspaper}
\begin{equation}\label{1_reradiated_field_unit_current}
\mathbf{E}^s(Z_L) = \mathbf{E}^s(Z_L=\infty)
-I_L\mathbf{E}_{I},
\end{equation}
where $\mathbf{E}_I$ is the radiated $E$-field in 
the transmitting (TX) mode for a
unit input current excitation, 
$\mathbf{E}^s(Z_L=\infty)$ is the 
scattered
$E$-field 
when its feed point is open-circuited, 
and $I_L$ is the load current.
The first and second terms on the right side in
(\ref{1_reradiated_field_unit_current}) are often
referred to as structural scattering and reradiation
components, respectively~\cite{tuley_knott1993_ch10,
collinspaper}. This decomposition of the total scattered
field is not unique, and the structural scattering
is set by the reference load condition. We choose
the open-circuit load as the reference. The total
scattering can be split in different
ways~\cite{hansen_ieeejproc1989,collinspaper}.

The scattered field decomposition
(\ref{1_reradiated_field_unit_current}) 
also applies to the general 
receiving
(RX) antenna array with
aperiodic loadings. To design a finite-dimension 
anomalous reflector,
a finite array variant of an isolated
antenna scattering~\cite{collinspaper} or
the infinite periodic array scattering~\cite{vuyyuru}
can be adopted.
In an $N$-element loaded meta-atom
array,
a complex load $Z_{Ln}$
$(n=1,\ldots,N)$
terminates the $n$-th
meta-atom.
While scattering synthesis via
(\ref{1_reradiated_field_unit_current}) can be
performed for any incident wave and array configuration, 
let us focus on plane-wave scattering by a linear
array.
In an $e^{j\omega t}$ time convention dependence, 
the finite structure is illuminated by an incident 
plane wave with an $E$-field 
\begin{align}
\mathbf{E}^i &= \hat{y}E^i_0e^{-jk(x\sin\theta^i+z\cos\theta^i)}
~\text{for TE polarization},\\
\mathbf{E}^i &= \hat{\theta}^iE^i_0e^{-jk(x\sin\theta^i+z\cos\theta^i)}
~\text{for TM polarization},
\end{align}
where $\theta^i$ is the angle of incidence, 
and $k$ is the free-space
wavenumber. Via reciprocity, the RX mode parameters are
related to
to the TX mode scanning in the direction 
$(\theta^s,\phi^s) = (\pi -\theta^i,\phi^i-\pi)$. 

In the RX and scattering 
case,
the open-circuit voltage and the load current across 
the terminals of the
$N$ loaded RX antenna array take the matrix form, 
as
an extension of the isolated RX antenna
case, written as
\begin{align}
\mathbf{V}_\text{oc} &= \mathbf{h}(\theta^s,\phi^s)
\cdot\mathbf{E}^i(O),
\label{Voc}\\
\mathbf{I}_L &=\left(\mathbf{Z}_A+\mathbf{Z}_L\right)^{-1}
\mathbf{V}_\text{oc},
\label{IL_port_current}
\end{align}
where $\mathbf{Z}_A$ is the $N\times N$ impedance 
matrix of the antenna array 
(a full matrix with elements representing
self and mutual impedances)
and $\mathbf{Z}_L$ is a diagonal matrix 
with 
$Z_{Ln}$ $(n=1,\ldots,N)$ along the diagonal.
The $N\times 1$ column vector 
$\mathbf{h}(\theta^s,\phi^s)$ has the vector 
effective height~\cite{collin1969}
for the $n$-th antenna as its $n$-th
element,
$\mathbf{h}_n(\theta^s,\phi^s)$,
computed from the TX mode 
for far-field observation
in $(\theta^s,\phi^s)$. 
The $N\times 1$ column vector $\mathbf{V}_\text{oc}$ 
contains all open-circuit voltages at the terminals 
of the RX array.
Its elements are
$V_{\text{oc},n}
=\mathbf{h}_n(\theta^s,\phi^s)\cdot\mathbf{E}^i(O)$
$(n=1,\ldots,N)$, where the incident 
field values $\mathbf{E}^i(O)$ are taken at 
the global coordinate origin $O$, as illustrated in
Fig.~\ref{fig:figure_theory}(a).
The column vector $\mathbf{I}_L$ contains
all individual load currents 
$I_{Ln}$ $(n=1,\ldots,N)$
at the array ports.
Canonical array configurations
with available near-zone field distribution
expressions allow analytical
load port current calculations,
such as for 2-D wire arrays~\cite{li2023tunable}.
Such an approach is 
not that straightforward for arbitrary  arrays,
because usually the elements lack
analytical formulations for near-zone
fields. In this case, load current calculation 
using the RX
antenna theory is useful.

It is essential to perform a set of preliminary 
simulation studies to prepare for global numerical 
optimization to attain the scattered beam in 
the anomalous direction by optimizing
$Z_{Ln}$'s.
The network
impedance matrix $\mathbf{Z}_A$
is found using a TX simulation for
mutual coupling between ports. TX simulations also provide
the vector effective height $\mathbf{h}_n$ 
for the $n$-th patch element.
To find $\mathbf{h}_n(\theta,\phi)$, 
it should be remembered to enforce 
$I_{Lm}=0$ $(m\neq n)$ by open-circuiting 
unexcited patch ports.
The open-circuit voltages for the $N$
element ports
can be computed using (\ref{Voc}).
Alternatively, $\mathbf{V}_\text{oc}$ can be
obtained
directly using an
RX-mode simulation
by
monitoring the voltage at the open-circuited ports.
Numerical studies confirm that the two approaches 
produce the same values of $\mathbf{V}_\text{oc}$.
Once the preliminary numerical analysis
is complete, optimizations of 
\emph{algebraic} nature
can be carried out to synthesize the desired
scattering characteristics. The required preliminary
analysis is unique to every array scattering
synthesis problem.

In the far-field region of the scattering
array,
the total secondary $E$-field, 
$\mathbf{E}^s(\mathbf{Z}_L)$, 
scattered by the array
for a particular load impedance matrix, $\mathbf{Z}_L$, 
in a reflection direction given by
$(\theta^r,\phi^r)$
is evaluated by extending 
(\ref{1_reradiated_field_unit_current}) as
\begin{equation}\label{radiated_E_field}
\mathbf{E}^s(\mathbf{Z}_L) = 
\mathbf{E}^s(\mathbf{Z}_L=\infty)
-\sum_{n=1}^N
\frac{jk\eta}{4\pi}
I_{Ln}\mathbf{h}_n(\theta^r,\phi^r)
\frac{e^{-jkr}}{r}
, \end{equation}
where $I_{Ln}$ and $\mathbf{h}_n(\theta^r,\phi^r)$
are
the load port current and the vector effective 
height, respectively, at the $n$-th port. 
The free-space intrinsic impedance is denoted
by
$\eta$, and
$\mathbf{E}^s(\mathbf{Z}_L=\infty)$ is the scattered field 
from the patch array
in $(\theta^r,\phi^r)$
when all
load terminals are open-circuited.
Let us limit our attention to anomalous reflection
in the $xz$-plane, i.e., $\phi=0$ or $\pi$.
We adopt the
bistatic scattering cross-section (SCS)
$\sigma$
as the metric in assessing the effectiveness
of the synthesized loaded array for creating a
strong scattered beam in the desired direction.
For the loaded array,
it is given by~\cite{balanis2015antenna}
\begin{equation}\label{SCS}
\mathbf{\sigma}(\mathbf{Z}_L,
\theta^r) = 
\lim_{r\to\infty}4\pi r^2
\frac{|\mathbf{E}^s(\mathbf{Z}_L)|^{2}}{|\mathbf{E}^i(O)|^{2}}.
\end{equation}
Once $\textbf{E}^s(\textbf{Z}_L=\infty)$,
$\mathbf{Z}_A$, and 
$\textbf{h}_n(\theta,\phi)$ are computed and available,
we note that $\sigma$ can be evaluated in every 
scattered direction
for any set of loads via an algebraic process.
It is numerically efficient
compared with full-wave simulations. Changing the
incident wave direction or polarization requires
new evaluation of $\textbf{E}^s(\textbf{Z}_L=\infty)$
only.

In addition to arrays of finite dimensions both
in the $x$- and $y$-directions, we will also consider
arrays that are finite along the $x$-axis
and infinite periodic along the $y$-axis
(i.e., finite~\texttimes~infinite).
For such arrays, anomalous reflection in the
$xz$-plane will be synthesized.
With a subwavelength periodicity
$b$ and a uniform phase
in the $y$-direction,
the scattered waves are cylindrical waves instead
of spherical waves. In the same manner as in
(\ref{radiated_E_field}), 
cylindrical wave scattering in the effective
2-D environment can be split into structural
and reradiation contributions
as
\begin{multline}
\mathbf{E}^s(\mathbf{Z}_L)=\mathbf{E}^s(\mathbf{Z}_L=\infty)\\
-\sum_{n=1}^N
\sqrt{\frac{k}{8\pi}}\frac{\eta e^{j\pi/4}}{b}
I_{Ln}\mathbf{h}_n(\theta^r)
\frac{e^{-jk\rho}}{\sqrt{\rho}},
\end{multline}
where $\rho=\sqrt{x^2+z^2}$ is the distance
from the $y$-axis. Then,
the 2-D scattering width, $\sigma_\text{2D}$,
is \cite{balanis1989}
\begin{equation}
\sigma_\text{2D}(\mathbf{Z}_L,\theta^r)=
\lim_{\rho\to\infty}2\pi\rho
\frac{|\mathbf{E}^s(\mathbf{Z}_L)|^2}
{|\mathbf{E}^i(O)|^2}.
\label{sigma_2d}
\end{equation}

\subsection{Optimization Methodology}\label{subsec:Opt_method}

We aim to achieve high-efficiency anomalous
reflection using reactive  loads only, without
any active elements.
Optimization of 
the
induced current distribution at 
the reactive loaded elements intends to redirect 
all the incident power from any oblique angle into 
any desired direction. Each reactive
load at the antenna terminals is independently varied,
resulting in an
algebraic optimization over the metasurface. 
A practical choice for the tunable components for 
the finite patch arrays is lumped elements 
(such as varactors or switches) 
in each unit cell. Another possibility is to use
electrically tunable substrate 
materials~\cite{IntMetaLiu2019}. 
The optimization process can be performed by 
maximizing $\sigma(\textbf{Z}_L)$ in the
desired direction by adjusting $Z_{Ln}$ $(n=1,\dots,N)$.
For a finite-array reflector
design, a bistatic SCS is an appropriate
quantity 
to maximize the scattering in the desired anomalous 
direction.

In preparation for the optimization, the
goal
is to maximize the predicted bistatic SCS
$\sigma$ in (\ref{SCS})
to realize perfect anomalous reflection in 
the desired anomalous
direction as the global fitness function. 
Numerical experiments reveal that
it is worth adding constraints designed to suppress
unwanted deflections in the specular reflection
and other directions
as a cost function
during reactive load optimizations.
These constraints 
define a
multi-objective optimization problem toward finding a globally optimal solution.
The multi-objective optimization problem is formulated as follows:
\begin{equation}\label{optimization_problem}
\begin{aligned}
&
& & \max_{\mathbf{Z}_L}
\mathbf{\sigma}(\mathbf{Z}_L
,\theta^r)\\
& \text{s.t.}
& & \mathbf{\sigma}
(\mathbf{Z}_L,\theta) \leq \epsilon,\\ 
& & &\forall\theta\in[-90\degree,\theta^r-\Delta\theta]
\cup[\theta^r+\Delta\theta,90\degree]\\
& \text{where}
& & \mathbf{Z}_L = \{ Z_{L1},Z_{L2}, \ldots, Z_{Ln} \}.\\
\end{aligned}
\end{equation}
Here, $\epsilon$ is a small positive constant.
The azimuth angle is limited to
$\phi^r=0$ or $\pi$, as appropriate for a reflection
in the $xz$-plane.
The optimization is performed using the 
\textit{fmincon} function in \textsc{Matlab}. 
The \textit{fmincon} function is a
practical tool for a multi-objective optimization.
The optimization converges to find an optimal 
set of
purely reactive load impedances using the 
objective function to maximize and the
additional constraints to minimize.


The optimization process is
as follows: First, an initial set of load impedances 
are selected
to initiate the
optimization. A set of
purely-reactive load values for 
the
$N$ meta-atoms
are
chosen for this purpose, e.g., 
from the traditional locally periodical approximation in the linear 
phase-gradient method. 
Details of the legacy conventional reflectarray 
method are given in Appendix~\ref{app:A}. 
Then, the bistatic SCS is calculated using (\ref{SCS})
over a range of $\theta$ as needed.
The above step is repeated by updating
the
purely reactive
impedance loads until the termination criterion 
is satisfied. Finally, we evaluate
the angular response in the far zone using 
the
optimized load values.
We verify whether the incident power is 
rerouted to the predetermined anomalous
direction, i.e., a near-perfect anomalous deflection is
achieved.


\section{Optimized finite  Arrays to realize Perfect Anomalous Reflectors}\label{sec:finite_Arrays_Opt}

It is expected that optimization of loads of 
$\lambda/2$-periodical arrays will not result in 
desired nearly perfect performance for any deflection angles, because it is known that realization of desired induced currents in general requires complex (active-lossy) loads \cite{li2023tunable}. For this reason, we
verify the efficiency of the optimization
and the performance of the optimized
reflectors
considering array designs with subwavelength unit cells 
and aperiodic passive loadings. 
A
subwavelength cell size enables a fine-grained
control of the induced current to help focus
the scattered wave
toward the desired 
direction~\cite{Smart_Radio_Environments}. 
The power
efficiency of anomalous reflection for
the conventional reflectarray design with a 
linear phase gradient loading diminishes 
gradually for large deflection tilts 
(considerably more than $50\degree$ at normal illumination) 
due to parasitic reflections into unwanted
directions~\cite{diaz2017generalized,vuyyuru}. 
For infinite periodic reflectors,
wide-angle anomalous reflector design techniques
that do not require supercell-level EM simulation
optimizations have been reported, based on
spatially continuous surface reactance 
synthesis~\cite{PlanarawplDohoon} and scattering
synthesis for infinite periodic loaded
arrays~\cite{vuyyuru}. However, these techniques
are limited to periodic reflectors.

\begin{figure}[t]
  \begin{center}
  \includegraphics[width=3.45in,height = 0.9in]{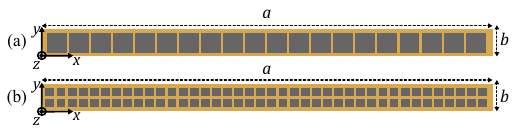}
  \caption{The schematic representation of the front view of
  a single $y$-period of
  the finite array in the $x$-direction
  with a dimension $a$, designed with a
  (a) half-wavelength 
  and (b) quarter-wavelength element spacing.
  The period along the $y$-axis direction is
  $b$.}
  \label{fig:figure_model}
  \end{center}
\end{figure}

In this section, we present finite-dimension
anomalous reflector designs
using arithmetic optimization of aperiodic loadings. 
First, we inspect
how many optimization parameters
are needed
and what
sub-wavelength element spacing is required to 
realize a highly 
directive non-specular reflection. We need 
the classical half-wavelength sampling 
from the Nyquist rate to control and construct 
the propagating EM spectrum under arbitrary illumination and scattering conditions. However, 
to fully control amplitudes and phases at all elements, complex-valued load impedances are required.
In studies of periodic arrays it was shown that to realize the currents 
that create desired scattering
using only passive elements,
twice as many scattering elements
are needed~\cite{popov2019Feb}. 
In this study, linear arrays of a
$\lambda/2$ spacing
and a $\lambda/4$ spacing are designed and compared, 
while the overall array size is fixed,
as illustrated in Fig.~\ref{fig:figure_model}.
This way, 
the impact of the element spacing on the
anomalous reflection characteristics can be inspected.
In addition, the performance of the anomalous reflectors is evaluated for both TE- and TM-polarized incident waves. 

We begin with determining the dimensions for 
the single meta-atom
as a self-resonant patch under
periodic boundary conditions. 
All numerical simulations have been
performed using a
commercial software 
(\textsc{CST Microwave Studio}). 
A linearly polarized square PEC patch antenna 
in a
unit cell is simulated at a design frequency of 28 GHz
to acquire a good impedance match to $50~\Omega$. 
The substrate used in all simulations is Rogers 
RO4350B (neglecting losses) laminate 
(a relative permittivity $\epsilon_r=3.66$) and a 0.338-mm thickness. The patch dimensions and feeding position data optimized for a
$50$-$\Omega$ input impedance at
resonance at the design frequency are reported in 
Table~\ref{tab:dimentions}. It is worth mentioning that the self-resonance and especially the $50$-$\Omega$ matching condition are used here as conventions, 
but
they are not necessary for a successful design.
Next,
$N$ individual aperiodic loads connected to the patch 
terminals are optimized following 
Section~\ref{sec:methodology} for 
$y$-periodic arrays.
The proposed design technique
can be used to synthesize large square or rectangular 
arrays toward practical applications.
However, the design
principle and effectiveness can be demonstrated
at a reasonable computational cost by
using 
$y$-periodic ($y$-infinite) arrays with one finite and one infinite dimension.
Here, the incidence and reflection is 
limited to the $xz$-plane.


\begin{table}[t]
\renewcommand{\arraystretch}{1.5}
    \begin{center}
    \caption{Dimensions of the $y$-periodic array}
    \label{tab:dimentions}
    \footnotesize
    \begin{tabular}{|p{1.5cm}|c|c|c|} \hline
        \multicolumn{2}{|c|}{ Parameters} & $\lambda/2$ spacing & $\lambda/4$ spacing \\ \hline   
        \multicolumn{2}{|c|}{\multirow{2}{*}{Array dimension $(a\times b)$}} & \multicolumn{2}{c|}{$10\lambda\times0.5\lambda$} \\ 
        \multicolumn{2}{|c|}{} & \multicolumn{2}{c|}{($107$~mm $\times~5.3534$~mm)} \\ \hline
        \multicolumn{2}{|c|}{Substrate height ($h$)} & \multicolumn{2}{c|}{$0.338$~mm } \\ \hline 
        \multicolumn{2}{|c|}{Square Patch dimension} & $2.554$~mm & $2.347$~mm \\
        \hline      
        \multirow{2}{=}{Probe position $(x_p,y_p)$} & TE Pol & $(0,0.444)$~mm  & $(0,1.068)$~mm \\ \cline{2-4}
        & TM Pol &$(0.444,0)$~mm  & $(1.068,0)$~mm \\ \hline
       \multicolumn{2}{|c|}{Total loaded elements $(N)$} &  $20$  & $80$ \\ \hline
    \end{tabular}
    \end{center}
\end{table}

\subsection{Optimized
Patch Arrays with a $\lambda/2$ Spacing}
\label{subsec:Sub-wavelength}

\begin{figure*}[t]
 \centering
 \includegraphics[width=7in,height = 2in]{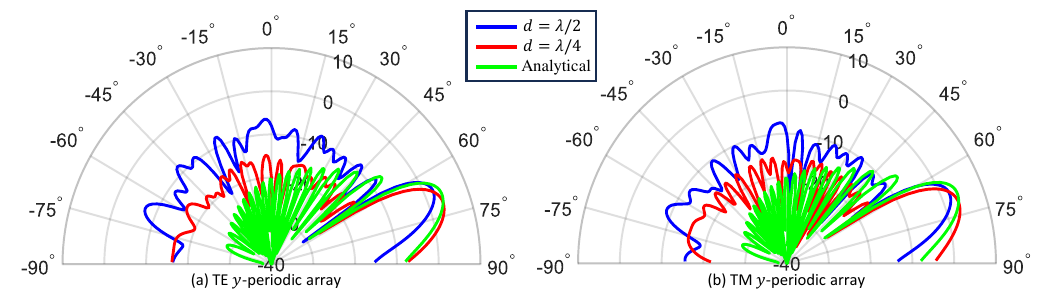}
 \caption{CST simulated $xz$-plane SCS patterns of the $y$-periodic anomalous reflector designs comparing $\lambda/2$ (blue solid line) and $\lambda/4$ (red solid line) element spacing results. These figures represent 1-D ($\phi =0$) scattering pattern cuts of the bistatic SCS. (a) $y$-periodic arrays for TE polarization and (b) for TM polarization. A theoretical bistatic SCS pattern (green solid line) for $y$-periodic arrays is plotted through analytical expressions for plane wave scattering from a 2-D strip for both (a) TE and (b) TM polarized fields, as derived from \cite{balanis1989}.}
 \label{fig:1D_half_quarter_lambda}
\end{figure*}

As discussed above, we start from a design of $\lambda/2$-spacing arrays as objects for comparison with subwavelength-spacing  arrays.
Two 
loaded patch arrays of a
dimension of $10\lambda\times0.5\lambda$ 
are
designed with a
half-wavelength element spacing comprising 
20 individual patch elements,
in TE
and also in TM 
polarization, 
as depicted in Fig.~\ref{fig:figure_model}(a). 
As an example, for an illuminating plane wave at
normal incidence $[(\theta^i,\phi^i)=(0,0)]$, 
anomalous reflection into
+70\textdegree{}
$[(\theta^r,\phi^r)=(70\degree,0)]$ is desired.
A starting point for the optimization is obtained
the purely-reactive load values 
from the conventional 
reflectarray method (Appendix~\ref{app:A}). The optimization
is carried out for maximizing
$\sigma_\text{2D}(\mathbf{Z}_L,\theta^r)$
for $y$-periodic arrays.

Figure~\ref{fig:1D_half_quarter_lambda} 
presents
the optimized performance for bistatic SCS
$\sigma$ for 
the 2-D scattering width $\sigma_\text{2D}$
in dBm (dB over meter)
for $y$-periodic arrays
in the $xz$-plane. 
The blue lines represent the scattering patterns for the half-wave spaced arrays.
The incident power is rerouted to
the desired +70\textdegree{} direction
using a
20-element finite-sized patch array
along
the $x$-dimension.
For an infinite periodic reflector,
a 0\textdegree-to-70\textdegree{} deflection requires
a supercell period of 
$\lambda/\sin70\degree\approx 1.0642\lambda$,
which
is not equal to any integer multiple
of the unit cell $x$-dimension for these finite arrays.


For the $y$-periodic arrays, the structure designed for TE polarization
(Fig.~\ref{fig:1D_half_quarter_lambda}(a)) gives a $-7.7$~dB  side lobe level (SLL) for the highest unwanted side-scattering direction compared to the desired scattering direction. Finally, an  SLL of $-9.2$ dB is achieved for TM polarized fields, as shown in Fig.~\ref{fig:1D_half_quarter_lambda}(b). Although the goal is to attain scattering towards a specific desired angle, the scattering peak is not aligned precisely with the intended angle. A beam pointing error of $3\degree$ for 
the
$70\degree$ $y$-periodic design is observed. This deviation arises from the constraints imposed by the finite number of individual patch-loaded elements and the relatively coarse half-wave patch spacing in the antenna array.

On top of the SLL, a green-colored SCS pattern 
is plotted for 
an ideal anomalous
reflection characteristic, which represents
a 100\% efficiency. It is associated with a reflection
coefficient with a linear phase gradient and a 
constant magnitude
corresponding to locally power-conserved reflection.
Physically, this ideal model is associated with
an active-lossy load impedance distribution. Under
the physical optics approximation, analytical results
for the 2-D SCS
are derived following \cite{balanis1989}.
Then, the anomalous reflection
efficiency is evaluated as the
aperture efficiency of scattering into the anomalous
direction. In other words, it
is defined as the ratio of
the 2-D SCS of the optimized physical design 
to the 2-D SCS of the
ideal, analytical model. 
The TE polarized results depicted in Fig.~\ref{fig:1D_half_quarter_lambda}(a) demonstrate an efficiency of 56.9\%. Similarly, TM polarized results in Fig.~\ref{fig:1D_half_quarter_lambda}(b) produce an efficiency of 63.2\% compared to the 2-D SCS analytical model.




\begin{figure}[t]
\centering
\includegraphics[scale=0.9]
{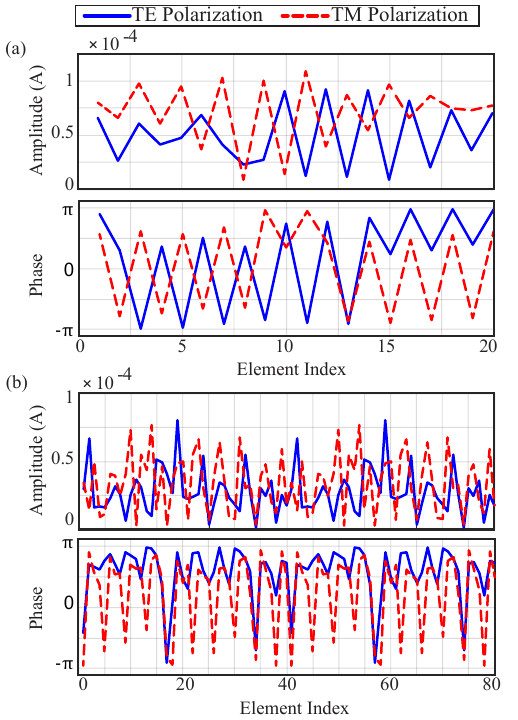}
\caption{The CST simulated amplitude and phase of the current at the patch antenna terminals to manipulate the reflected wavefronts
for TE (blue solid line) and 
TM (red dashed line) polarizations. 
(a) $y$-periodic arrays with a
$\lambda/2$ element spacing.
(b) $y$-periodic arrays with a 
$\lambda/4$ element spacing.}
\label{fig:port_current_half}
\end{figure}

\subsection{Optimized Patch Arrays
with a $\lambda/4$ Spacing}
\label{subsec:Quarter-wavelength} 

To provide a fair comparison with the half-wave spaced array, 
finite$\times$infinite, $y$-periodic
arrays with a periodic cell of
size $10\lambda\times 0.5\lambda$ 
comprising 80 individual patches
for both TE and TM polarizations with a
$d=\lambda/4$ spacing are
designed, as depicted in Fig.~\ref{fig:figure_model}(b). 
We have now 40 independent load reactance variables, since the two loads on each column should have identical values in order to keep the scattering
in the
$xz$-plane. Repeating  the same procedure as in the half-wavelength-spaced design,  first, we obtain the initial set of purely-reactive load impedances from the conventional reflectarray method (Appendix A), and then an  optimization is performed using the analytical expression for
2-D bistatic SCS (\ref{sigma_2d}) for $y$-periodic arrays.

The red curves in Fig.~\ref{fig:1D_half_quarter_lambda} 
depict the results for $\lambda/4$-spaced
$y$-periodic array.
In
TE polarization, the ratio of the bistatic SCS 
for the desired lobe to the highest side lobe 
gives $18.2$~dB, as seen in 
Fig.~\ref{fig:1D_half_quarter_lambda}(a). 
Likewise, the results for TM polarized fields 
are presented in Fig.~\ref{fig:1D_half_quarter_lambda}(b), 
showing the highest
SLL of $-18.6$~dB. The $\lambda/4$ periodicity design performance is much better as compared
with
the $\lambda/2$ results. 
This is an expected result, as the array density is doubled in both dimensions compared
with
the half-wave spaced case, giving a possibility 
to control 
the
subwavelength current distribution at the reflector array, which is needed to remove the need of active-lossy loads of the array elements.
As depicted in Fig.~\ref{fig:1D_half_quarter_lambda}(a), 
the TE polarization reports an efficiency of 
92\%, while the TM polarized results in 
Fig.~\ref{fig:1D_half_quarter_lambda}(b) exhibit
a similar
efficiency of 93.8\% compared to the 2-D SCS 
analytical model. The scattering beam peak for 
the $\lambda/4$ spacing aligned closer
with the intended desired angle compared to $\lambda/2$ 
spacing.
Here, we observe that the reflection
efficiencies of the optimized arrays do not closely
approach 100\%. This is because we are requiring
minor scattering lobe levels to be low, as described
in Section~\ref{subsec:Opt_method}. This
is analogous
to targeting low SLLs
in antenna synthesis. A tradeoff relation 
between the SLL and the main beamwidth leads to a
slightly weaker SCS in the anomalous direction
when minor scattering lobes are suppressed.
Specifically,
the ideal anomalous reflection is associated with
a uniform aperture field for the scattered wave,
giving a scattering SLL of \textminus10.5~dB.
This is higher than the scattering SLL values of
the optimized arrays.
 

The $\lambda/4$ spacing results show 
excellent performance for 
the
bistatic SCS
with the optimized distribution of reactive loads. The simulated outcomes reveal a dominant $70\degree$ scattering lobe while  
other scattering lobes are suppressed, 
demonstrating a nearly perfect angle-tunable anomalous deflection. 
In contrast to
the
$\lambda/2$-spaced arrays, a
$\lambda/4$-spaced design 
offers more degrees of freedom, resulting in better optimization.
The induced currents flowing 
through
the patch antenna terminals to anomalously deflect 
an incident plane wavefront into the desired 
direction with a fixed $\lambda/2$ and $\lambda/4$ 
inter-element spacing are presented in 
Fig.~\ref{fig:port_current_half},
and they
shows aperiodic distribution 
behaviors.


Furthermore, it is noteworthy to discuss the 
optimization speed to 
assess the efficiency
of the antenna scattering synthesis method. 
Once the pre-computed data are
acquired from the TX antenna simulations, the optimization time 
increases
with an increased number of
loaded elements.
The time duration for $20$ and $40$ individual
load
optimization for $\lambda/2$ and $\lambda/4$ spacing, respectively, is less than a minute to obtain 
the optimum reactive loads. 
We utilize an HP EliteBook 845 G7 notebook PC 
with Windows 10 Enterprise as our hardware platform.
Thus, this technique offers scattering maximization in the desired direction by avoiding computationally intensive full-wave EM simulations and prolonged optimization times.


\begin{figure}[t]
  \centering
  \includegraphics[width=3.4in,height = 2.3in]{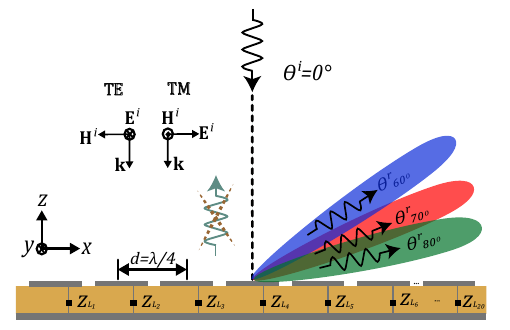}
  \caption{Side view of the planar array under a 
  normally
  incident plane-wave illumination and the desired large deflection angles
  for
  TE and TM
  polarizations.}
  \label{fig:figure_planar_model}
\end{figure}

\section{Finite Planar Antenna Array}\label{sec:Planar_Square_Antenna}

After confirming that the design methodology described in Section~\ref{sec:methodology} works well for semi-finite arrays, and that a quarter-wave spacing between unit cells is, in terms of scattering performance, superior to 
a
half-wave spacing, 
we extend the study to finite-size planar
arrays, as depicted in Fig.~\ref{fig:figure_planar_model} with $5\lambda\times 5\lambda$ physical 
dimensions. With a quarter-wavelength spacing, the array
comprises $400$ individual loads as free optimization parameters for both TE and TM
polarizations.
Let us emphasize that we can use the reactive impedances of all the loads as independent optimization variables, as needed when anomalous
reflection occurs in a direction
off the plane of incidence.

\begin{figure*}[b]
\begin{center}
\includegraphics[width=7in,height = 2in]{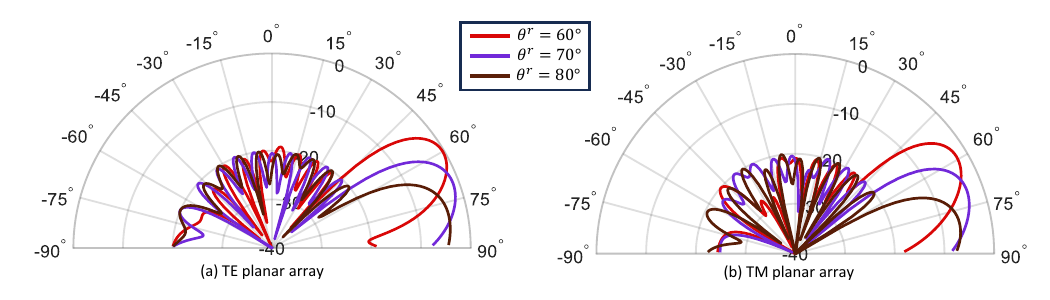}
\caption{The CST simulated SCS patterns 
in the $xz$-plane
of the anomalous reflectors with 
an
overall dimension of $5\lambda\times 5\lambda$ 
with 
a
$\lambda/4$ element spacing.
(a) TE polarization. (b) TM polarization.
}
\label{fig:2D_Prototype_res}
\end{center}
\end{figure*}

Figure~\ref{fig:figure_planar_model} illustrates 
deflecting the normally 
incident plane wave into 
$60\degree$, $70\degree$, and $80\degree$ 
by the
finite array designed for TE and TM
polarizations.
Deflection angles less than 
60\textdegree{} are omitted
because
it is assumed that they can be realized 
efficiently  using the conventional linear phase 
gradient method. A large deflection of $\theta^i = 0\degree$ to $\theta^r = 60\degree$, $70\degree$, and $80\degree$ propagating angles for an infinite periodic reflector requires a supercell period of $\lambda/\sin\theta^r\approx 1.1547\lambda$, $1.0642\lambda$, and $1.0154\lambda$, respectively. These periodicities do not synchronize with the fixed array period, but continuous scanning across all 
anomalous reflected beams 
is enabled
using single-period planar finite arrays. 

Again, we follow the procedure of Section~\ref{sec:methodology}, determining $\mathbf{Z}_A$, $\mathbf{h}(\theta^s)$ and  $\textbf{E}^s(\textbf{Z}_L=\infty)$ from full-wave EM simulations using CST.
Then, the loads are optimized to steer the incident wave into a desired reflection angle, maximizing $\sigma(\mathbf{Z}_L,\theta^r)$.


\begin{figure}[t]
 \centering
\includegraphics[width=3.45in,height = 1.8in]
{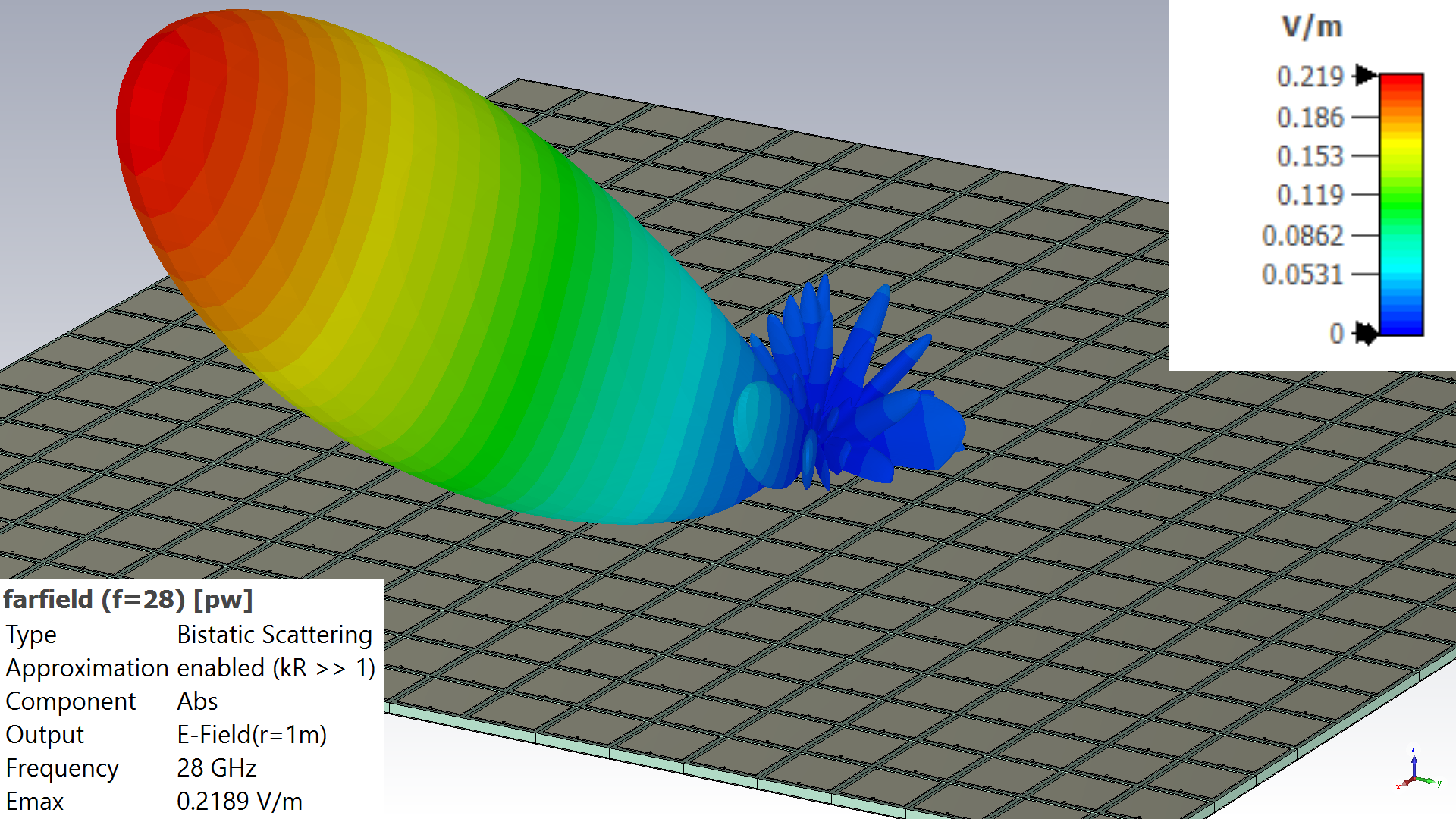}\\
\caption{CST simulated 3-D 
scattering pattern
of the 0\textdegree-to-60\textdegree
anomalous reflector design with 
an
overall dimension of $5\lambda\times 5\lambda$ 
with 
a
$\lambda/4$ element spacing 
in TM polarization.
}
\label{fig:3D_Epattern}
\end{figure}

Figure~\ref{fig:2D_Prototype_res} 
plots the predicted SCS patterns of 
the optimized planar array with 
a
$\lambda/4$ element spacing.
The results have been confirmed by separate
full-wave scattering simulations of the optimized
designs.
The reported outcomes show a dominant scattering beam into the desired direction while 
minimizing
the specular lobe and other undesired reflections. For the TM polarized case, the SLL of the strongest
minor scattering beam  
compared with the field at the desired direction reports $-18.6$~dB, 
$-16.6$~dB, and $-13.5$~dB for the deflection angles 
$60\degree$, $70\degree$, and $80\degree$, 
respectively. Likewise, the designs for TE polarized fields  
have SLLs
of $-18.8$~dB, $-18.6$~dB, and $-16.2$~dB for the 
respective reflection angles,
which are
similar to the SLLs of the 
TM results. The achievable optimized scattering cross-section inevitably decays when the deflection angle approaches 90\textdegree{}, 
as the projected area of the scattering surface seen in that direction diminishes.
Figure~\ref{fig:3D_Epattern} demonstrates the 
3-D magnitude plot of the 
scattered $E$-field of the anomalous 
reflector design for the TM 
polarization, deflecting a normally
incident plane wave into $60\degree$ with 
minimization of all undesired side lobes.
In addition to the narrow half-power beamwidth 
of 19.2\textdegree{} in the
$E$-plane owing to collimation via anomalous
reflection, the beamwidth in the
$H$-plane is narrow at 
9.2\textdegree.

The optimization
for $400$ individual load variables takes approximately $2$ hours to attain the optimal load 
impedances. 

\section{Experimental Results}\label{sec:RISExperimental}

To further validate the performance of the optimized reflectarrays for various deflection angles, we design, 
fabricate,
and characterize a compact  finite $28$-GHz anomalous reflector. 
From the $y$-periodic array design analysis in Section~\ref{sec:finite_Arrays_Opt}, 
a
$\lambda/4$ spacing is
chosen.
The available in-house PCB manufacturing facility and anechoic room test setup are taken into consideration in design to enable practical comparison of the CST simulation results with the experimental scattering pattern. A dual-layer printed circuit board 
(PCB) 
is selected
for manufacturing using an LPKF Proto laser and galvanic through-hole plating. 
For
measurements, an EM shielded anechoic room with dimensions $5\times5\times2.2$~m
is used.

\subsection{Practical Design and Manufacturing}\label{subsec:RISFabrication}

A finite planar array prototype is designed to be 
illuminated 
by a TM-polarized
plane wave 
at normal incidence
$[(\theta^i,\phi^i)=(0,0)]$ to have 
scattering
steered to a fixed
+70\textdegree{} $[(\theta^r,\phi^r)=(70\degree,0)]$  
anomalous reflection angle at 28~GHz.
The designed array has $1600$ 
metal patches 
arranged periodically with a
$\lambda/4$ period
in a $40\times40$ square array,
resulting in $10\lambda\times10\lambda$  
overall dimensions. 

The unit cell is a 
copper square patch on a PCB, 
using a
Rogers RO4350B laminate with LoPro 
copper layers ($\epsilon_r=3.55$, $\tan\delta =0.0037$,
a
$0.52$-mm thickness). For simulations, we use the lossy materials definitions available on CST's material library, assuming annealed copper with
a 0.035-mm thickness and surface roughness 
of 0.001~mm.
The square patch length
is 2.039~mm, and 
the feeding position $(x_p,y_p) = (1,0)$~mm 
is optimized for resonance
with a
$75~\Omega$ input impedance 
at the desired frequency. 
This is a practical choice, suitable 
for implementing
a set of desired load impedances with shorted and open coplanar waveguide (CPW) strips. 

Considering further practical implementations, using a wide range of continuously tunable load values might become infeasible. Consequently, the number of possible load values is limited to 
a
3-bit 
(a 45\textdegree{} increment) 
quantization resolution to have $8$ distinct phase responses of homogeneous arrays. This discretization of the corresponding load reactances will provide an
insight into
the robustness of the design method, showing whether inaccurate load impedance values can still provide good anomalous reflection. Additionally, 
optimization of the load impedances
converges
significantly faster with a limited number of 
available load reactance values.

\begin{figure}[t]
  \begin{center}
  \includegraphics[width=3.48in,height = 1.5in]{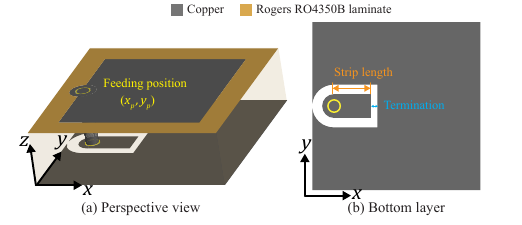}
  \caption{(a) Perspective view and 
  (b) bottom view
  of the proposed manufacturable single patch element, featuring 
  a terminated
  CPW line.}
  \label{fig:single_element_CPW}
  \end{center}
\end{figure}

We start our analysis with a
$10\lambda\times0.5\lambda$ $y$-periodic array 
incorporating 80 individual patch
elements.
The scattering behavior of 
a
$y$-periodic semi-finite array should be consistent with 
a large finite
rectangular array. 
The reactive loads are optimized following Section~\ref{sec:methodology}.
Then, we form a $10\lambda\times10\lambda$ panel with $1600$ elements as a set of $20$ optimized linear arrays,
to avoid
a large
preliminary set of simulations with $1600$ elements. As a final design step, 
a
mapping of the 3-bit load reactances into a practical, equivalent set of shorted and open CPW lines is made, as reported in Table~\ref{tab:CPWdimensions}. The geometry of the proposed single patch antenna, featuring 
a terminated CPW strip, 
is illustrated in Fig.~\ref{fig:single_element_CPW}.

\subsection{Measurements}\label{subsec:RISMeasurements}

\begin{table}[t]
\renewcommand{\arraystretch}{1.25}
    \begin{center}
    \caption{CPW strip dimensions}
    \label{tab:CPWdimensions}
    \footnotesize
    \begin{tabular}{|c|c|c|c|c|} \hline
    No. & Phase $(\varphi_{x}^r)\degree$ & Reactance $(\Omega)$ & Strip (mm) & Termination \\   \hline
    $0$ & 0 & $123$ & $1.156$ & short \\ \hline
    $1$ & $\pi/4$ & $62$ & $0.925$ & short \\ \hline
    $2$ & $\pi/2$ & $21$ & $0.635$ & short \\ \hline
    $3$ & $3\pi/4$ & $-13$ & $1.745$ & open \\ \hline
    $4$ & $\pi$ & $-52$ & $1.365$ & open \\ \hline
    $5$ & $-3\pi/4$ & $-125$ & $0.994$ & open \\ \hline
    $6$ & $-\pi/2$ & $-570$ & $0.6$ & open \\ \hline
    $7$ & $-\pi/4$ & $334$ & $1.4$ & short \\ \hline
    \end{tabular}
    \end{center}
\end{table}

\begin{figure*}[t]
\begin{center}
\includegraphics[width=7in,height = 2.25in]{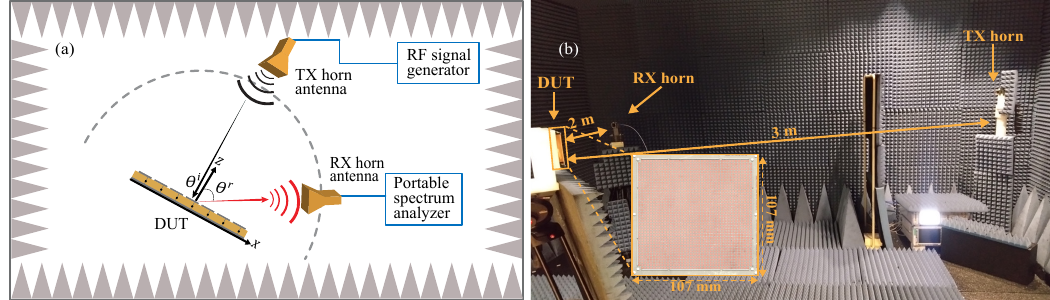}
\caption{Manufactured sample and measurement setup. 
(a) Illustration of the experimental setup for 
the fixed-angle anomalous reflector prototype (top view).
(b) A phogotraph
of the prototype and
measurement arrangement 
for
the scattering pattern in an
anechoic chamber.
}\label{fig:measurement_setup}
\end{center}
\end{figure*}

The manufactured anomalous reflector sample is tested in an anechoic chamber organized to carry out bistatic SCS measurements with the device under test (DUT) mounted on a stationary tripod. The physical size of the manufactured PCB panel is approximately $107$~mm ($x$-axis) by  $107$~mm ($y$-axis), and a 
photograph along with an illustration
of the experimental setup is shown in Fig.~\ref{fig:measurement_setup}.
Standard gain horn antennas with 
a
$18.5$~dBi gain at 28~GHz are 
used as 
TX
and 
RX
antennas. At a
3~m
distance from the DUT, the 
TX
horn antenna connected to a signal generator (SG) is 
mounted on a fixed pole. The 
RX
horn antenna is at 
a 2~m
distance from the DUT, with a portable spectrum analyzer (SA) as a detector. 

Although a proper far-field condition 
(the Fraunhofer distance 2.14~m) cannot be satisfied 
in the test setup, a decent estimate of 
the scattering pattern is obtained. 
In the current experimental setup, measurements are 
conducted at 28~GHz with the receiver moving at
a
2.5\textdegree{} angle resolution across the azimuth 
over $\theta^r\in[-90\degree,90\degree]$.
In order to facilitate measurement also 
in 
the specular reflection direction,
a 2\textdegree{}
downtilt is utilized, so that the height of the TX antenna is 
set to 140~cm, the height of the DUT to 129.5~cm, 
and the height of the RX antenna to 122.5~cm. The consequence of this positioning is that the angle of incidence for the DUT is not exactly broadside, 
but it is $(\theta^i,\phi^i)=
(2\degree,90\degree)$. 
The DUT then should reflect the signal to the designed anomalous azimuth angle $70\degree$ at a
$-2\degree$ elevation.

\begin{figure}[t]
  \begin{center}
  \includegraphics[width=3.45in,height = 1.8in]{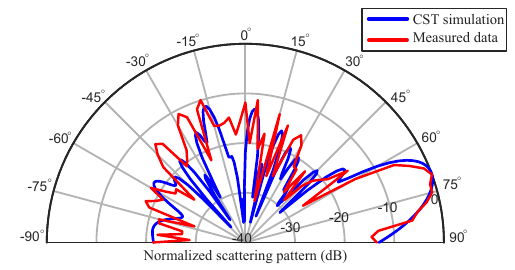}
  \caption{Comparison of the
  measured and simulated 
  results for the normalized
  scattering pattern of the anomalous reflector.}
  \label{fig:measurement_results}
  \end{center}
\end{figure}

Figure~\ref{fig:measurement_results} 
compares the normalized scattered patterns
obtained from the measurements and simulations,
where a
good agreement is observed for the whole angular range.
The red curve representing the measured data 
shows a \textminus10~dB
ratio of the bistatic SCS of the highest
side lobe to
the desired anomalous scattering lobe.

\section{Conclusion}\label{sec:conclusion}

A numerically efficient design strategy for planar finite-sized anomalous reflectors capable of scanning the deflection beam to any arbitrary angle without parasitic scattering 
has been demonstrated. 
Steering and
maximizing the directive scattered beam is cast 
to a
circuit domain-based algebraic optimization, 
which is numerically efficient compared
with
the traditional full-wave EM simulation-based 
techniques. 
A reflector comprises a loaded
array of
antenna elements treated as a linear multi-port
network. Reactive load impedances for individual
antennas are optimized to enhance the main scattering
lobe and suppress scattering in other directions.
The design approach can 
also
synthesize high-efficiency
wide-angle reflections using quantized reactive
loads, which may be useful in practical tunable
reflectors employing digital phase shifters.

This study was performed
using
a practical array of 
printed microstrip
patches 
for 
two- and three-dimensional models. 
The fabrication
and subsequent experimental validation
confirms
a large fixed-response
wide-angle reflector with
an optimized 
set of quantized reactive load values.
The results provide an
empirical evidence of effective design of anomalous reflectors realizing nearly perfect reflection into an arbitrary direction in a fixed-period configuration.
Although 
the proof of concept has been developed and 
validated for specific 
antenna structures at 
a
mmWave frequency,
the design approach is general 
and applicable to a wide range of reflectors
comprising loaded scattering elements,
as the scattered field is formulated in terms of the
generic radiation property of the array.
In a load optimization for
real-time reconfiguration applications, the impact of
losses in practical phase shifters can be incorporated
into the design method by introducing a resistance 
value to every load impedance.

Finally, this technique 
can be used to shape wave reflections 
in rather general ways, not limited to anomalous 
reflectors considered here. 
For example, there is a problem of  parasitic 
specular reflections and other scattering from 
offset fed reflectarrays, e.g. \cite{huang2008,Budhu}. 
Since
the developed method is based on a global optimization 
of array loads, it is possible to define the incident 
field as the field of a primary horn at an arbitrary 
offset position and find the optimal loads for 
specular reflection elimination,
while forming a main beam in the desired
direction via coherent
addition of anomalously reflected waves.


\appendices
\section{Conventional Reflectarray Method}\label{app:A}

We use the
traditional phase-gradient reflectarray design 
based on the
LPA as the initial guess for the
load reactance values and also as a reference.
Under the periodic boundary conditions,
the patch element in the subwavelength
unit cell is simulated for the plane-wave
reflection coefficient as a function of the
load reactance. This way,
we build a design curve as a mapping between 
the reflection phase and the load reactance.
Next, the local
reflection phase distribution
is calculated analytically from 
the linear phase gradient relation
associated with a given set of incidence and
reflection angles.
For a normally incident plane wave,
the reflection phase for $n$-th 
patch element 
with the patch center located at
$(x,y)=(nd,0)$
for a deflection angle $\theta^r$
is then given by

\begin{equation}
\varphi_{x}^r = e^{-jnkd\sin\theta^r}.
\label{phir_linphase}
\end{equation}

Next, we create a finite patch array 
loaded with the
set of reactive load impedances
associated with (\ref{phir_linphase})
to provide the
$x$-linear reflection phase gradient 
needed for
the anomalous deflection angle, read off of 
the design curve. 

\ifCLASSOPTIONcaptionsoff
  \newpage
\fi

\bibliographystyle{IEEEtran}

\bibliography{IEEEabrv,Bibliography}


\end{document}